# High-altitude signatures of ionospheric density depletions caused by field-aligned currents

T. Karlsson[1], N. Brenning[1], O. Marghitu[2,3], G. Marklund[1], S. Buchert[4]

[1] *Space and Plasma Physics, Royal Institute of Technology (KTH), Stockholm, Sweden.*
[2] *Institute for Space Sciences, Bucharest, Romania.*
[3] *Also at Max-Planck-Institut für extraterrestrische Physik, Garching, Germany.*
[4] *Swedish Institute of Space Physics, Uppsala, Sweden.*

**Abstract.** We present Cluster measurements of large electric fields correlated with intense downward field-aligned currents, and show that the data can be reproduced by a simple model of ionospheric plasma depletion caused by the currents. This type of magnetosphere-ionosphere interaction may be important when considering the mapping between these two regions of space.

## 1. Introduction

A system of magnetic field-aligned current sheets closing via Pedersen currents in the ionosphere will set up an ionospheric electric field. For constant conductivity, and for sheets extending to infinity along the field-line and one of the perpendicular directions, we get:

$$E_\nu = \frac{J_P}{\Sigma_P} = \frac{1}{\Sigma_P}\int j_\| dv = \frac{1}{\mu_0 \Sigma_P}\int \frac{\partial B_\tau}{\partial v} dv = \frac{1}{\mu_0 \Sigma_P} B_\tau \quad (1)$$

where $\nu$ is the direction perpendicular to the sheet, $\tau$ the tangential direction, $E_\nu$ is the normal electric field, $J_P$ and $\Sigma_P$, the height integrated Pedersen current and conductivity, $B_\tau$ the tangential magnetic field, $j_{//}$ the field-aligned current (positive for downward currents) and $\mu_0$ the magnetic permeability of vacuum. This kind of correlation between $E_\nu$ and $B_\tau$ can be seen rather often in the dayside auroral oval (e.g. Ishii et al., 1992). When the conductivity is not constant, the above correlation breaks down; in this paper we will present data from the Cluster spacecrafts, where this correlation is replaced with a correlation between $E_\nu$ and $j_{//}$, i.e. the derivative of $B_\tau$.

## 2. Cluster data

We present electric and magnetic field data from the EFW (Gustafsson et al., 1997) and FGM (Balogh et al., 2001) instruments on the Cluster satellites, which have an apogee of 19.8 $R_E$ and a perigee of 4.0 $R_E$, in radial distance. We first present data from a northern hemisphere auroral oval crossing, on Feb 18, 2004, from 08:58:20 to 09:10:00 UT. The Cluster radial distance during this time period was about 4.2 $R_E$, and the satellite separations between approximately 350 and 1100 km. In Figure 1 we show the residual magnetic field vectors along the satellite tracks projected onto a plane perpendicular to the geomagnetic field. The two perpendicular directions in the figure roughly correspond to geomagnetic North, and East. The diamonds at the bottom end of the tracks indicate the satellite positions at 08:58:20 UT. (The data is color coded: black – S/C 1, red – S/C 2, green – S/C 3, blue – S/C 4.) The satellites move relatively close to a pearls-on-a-string configuration. The main feature of the data is the crossing of three sheets of field-aligned current, from bottom to top a relatively smooth sheet of upward current approximately 800 km wide, a thinner sheet of downward current (≈250 km), and finally a wider sheet of predominantly upward currents (~1000 km wide). (The meridional mapping factor to ionospheric altitude is 11.6.) This current system remains essentially stationary in space for the whole 200 s period between the crossings of the central current sheet by S/C 1 and S/C 4, which is the reason we have chosen to present this event. We have applied minimum variance analysis on the magnetic field data from all four S/C, and have used the average resulting angle of 5.8° to establish the sheet-aligned coordinate system. We have then used the infinite current sheet approximation to calculate the field-aligned current $j_{//}$ from the tangential component of the residual magnetic field $B_\tau$. In Figure 2 we present $j_{//}$ and the normal electric field $E_\nu$ measured by Cluster. All values are mapped to ionospheric altitude. Also presented is the result of a model calculation described in Section 3. The correlation between $E_\nu$ and $j_{//}$ is clear for all S/C in the downward current region. This type of correlation is rather uncommon, but a manual inspection of around 300 auroral zone crossings resulted in identification of 23 similar events, i.e. in about 8% of the crossings, all for downward currents. 17 of the 23 events where encountered during winter conditions and 15 on the night side.

**3. Comparison data – model**

The close relation between the electric field and the local downward field-aligned current (DFAC) suggests that there is a relation between the DFAC and the conductivity, since an infinitesimally thin current sheet gives a negligible contribution to the ionospheric closure current across the sheet, $J_\nu$. However, with a coupling to a local decrease in the conductivity it can produce a local increase in $E_\nu$ (Figure 3). Such decreases in the conductivity coupled to DFACs have been modeled by Doe at al. (1995), Blixt and Brekke (1996), Karlsson and Marklund (1998, 2005), and Streltsov and Marklund (2006). A few radar observations of ionospheric density cavities which may be related to this mechanism have been reported by Doe et al. (1993), Aikio et al. (2002), and Nilsson et al. (2005). The reason that a cavity is formed in DFAC regions is that the parallel current is mainly carried by electrons, whereas the Pedersen current is carried by ions. In regions where the downward parallel and perpendicular currents couple there will then be a net outflow of current carriers.

Here we model this interaction in a heuristic way by prescribing the conductances by

$$\Sigma_P = \Sigma_{P,0} - \begin{cases} k_{down,s} j_{//}, & \text{for downward } j_{//} \\ 0, & \text{for } up\text{ward } j_{//} \end{cases}$$

$$\Sigma_H = 2\Sigma_P$$

(2)

where $\Sigma_{P,0}$ and $k_{down,s}$ (>0) are constants, with $s$ = 1-4, for the four spacecraft crossings. We ignore any effects on the conductance from the upward currents, since we will concentrate on the electric field behavior in the downward current region. We also set a minimum value for the Pedersen conductivity of 0.2 S, which represents the background conductivity due to galactic cosmic rays, which are always present. Current continuity and the assumption of an infinite current sheet yields

$$J_\nu(\nu) = \int_{\nu_0}^{\nu} j_{\parallel}(\nu')d\nu' + \Sigma_P(\nu_0)E_\nu(\nu_0) - \Sigma_H(\nu_0)E_\tau(\nu_0) \qquad (3)$$

where $J_\nu$ is the height-integrated ionospheric current normal to the sheet. $E_\tau$ is constant if we use the electrostatic approximation ($\nabla \times \mathbf{E} \equiv 0$). (2) and (3) then give

$$E_\nu(\nu) = \frac{\Sigma_{P,0}}{\Sigma_P(\nu)}E_\nu(\nu_0) + \frac{\Sigma_H(\nu) - \Sigma_H(\nu_0)}{\Sigma_P(\nu)}E_\tau + \frac{1}{\Sigma_P(\nu)}\int_{\nu_0}^{\nu} j_{\parallel}(\nu')d\nu' \qquad (4)$$

Using the observed values for $j_{//}$ along each of the satellite tracks, we can calculate $E_\nu$, as a function of $\Sigma_{P,0} = \Sigma_P(\nu_0)$, $E_\nu(\nu_0)$, $E_\tau$, and $k_{down,s}$. Before $t \approx 320$ s the electric field is small and rather constant and we can assume that it can be mapped to the ionosphere and be taken as the background field of our model. However, there is an offset in the electric field component aligned with the direction towards the sun, due to a photo electron sheet. Using data from the electron drift instrument (EDI) on S/C 1 we correct for this and then take the average electric field for 60 s prior to the crossing of the large DFAC, which we use as our background ionospheric electric field: $E_\nu(\nu_0) = 0$, $E_\tau = -6$ mV/m (values mapped to the ionosphere).

In principle the conductance could be calculated from the electron data, but this is a very uncertain procedure in the absence of energetic precipitating electrons, and outside the scope of this paper. Instead we assume a reasonable background conductance. The results are rather robust with respect to the chosen value of $\Sigma_{P,0}$, but the numerical value of $k_{down}$ will of course vary within a factor of 2-3 depending on the choice of conductance. By trial and errorr we then find that the following parameters reproduce the electric field behavior in the DFAC region well:, $\Sigma_{P,0} = 5$ S and $k_{down,1} = 0.33$ Sm$^2$/µA, $k_{down,2} = 0.43$ Sm$^2$/µA, $k_{down,3} = 0.44$ Sm$^2$/µA, $k_{down,4} = 0.68$ Sm$^2$/µA, where the subscript on the $k$'s indicate S/C number. $E_\nu$ thus calculated is plotted in green in Figure 2. Thus the same set of parameters, except for $k_{down}$, reproduces the DFAC electric field quite well. It is interesting that $k_{down}$ has an increasing trend with time; in Figure 4 we plot the values of $k_{down}$ as a function of time from the first crossing of the current sheet. The crossing time is defined as the time when the current maximum is encountered, and the error bars in the t-direction indicate when the current is half the maximum value. A linear fit is reasonable which means that we can write $k_{down} = \kappa \cdot (t - t_0)$, with $\kappa = 1.4 \cdot 10^{-3}$ Sm$^2$/µAs, and $t_0 \approx -200$ s, consistent with a gradual deepening of the density cavity, beginning about 200 s before the first satellite crossing.

Revisiting the data from the simulations by Karlsson et al. [1998] we can calculate $\kappa$. In the simulations, the development of $k_{down}$ settles down to a reasonably linear dependence on time after the first tens of seconds, from which we can estimate $\kappa$. The value of $\kappa$ depends on various initial conditions of the simulations but for some realistic situations varied from around $1 \cdot 10^{-5}$ to $2 \cdot 10^{-3}$ Sm$^2$/µAs, which is in agreement with the above measurement.

For this event the horizontal ionospheric current $J_{v,//}$, resulting from the feeding field-aligned currents was comparable to the current associated with the background electric field: $|J_{v,//}| \approx 20$ mA/m, $|E_\tau \Sigma_H(v_0)| \approx 30$ mA/m. Below we show two cases where one of these current contributions dominates over the other one.

First (Figure 5a) we show data from a northern hemisphere auroral oval pass on Apr 27, 2002, with MLT $\approx$ 22, ILat $\approx$ 66º, and the geocentric distance 4.9 $R_E$. We show data only from S/C 4, but similar signatures can be seen on S/C 2 and 3. Using the same method as above we calculate $j_{//}$, $\Sigma_P$, and $E_v$. In the figure the modeled $E_v$ and $\Sigma_P$ is plotted in red. $j_{//}$ is not shown, but has a maximum (downward) value of 34 µA/m$^2$. For this event upward accelerated electrons are observed from $t \approx 70950$ s, which complicates the mapping of the background ionospheric electric field. We instead here consider it as a free parameter. The fact that the constant background current (driven by the background electric field) dominates over $J_{v,//}$ (440 mA/m vs. 90 mA/m) means that the electric field traces out the form of the conductivity, which in turn traces out the DFAC. We thus get a very detailed correlation between the electric field and the DFAC, and a unipolar $E_v$ field signature at the density cavity.

In Figure 5b we present data from an auroral crossing on Jan 11, 2005. MLT $\approx$ 22, ILat $\approx$ 66º, geocentric distance 4.3 $R_E$, max($j_{//,down}$) = 32 µA/m$^2$. Here the background ionospheric current is dominated by $J_{v,//}$. This means that the ionospheric current is not constant across the low-conductivity region, and we should not expect such a detailed correlation between $E_v$ and $j_{//}$ as in the above case. In fact, what we see is that the electric field is large inside the low-conductivity region of the DFAC, but since the ionospheric current changes sign inside this region, the electric field also does, and produces a bipolar electric field signature. A small westward background electric field shifts the zero crossing of the total current $J_v$ slightly from that of $J_{v,//}$.

**4. Discussion and conclusions**

The correlation between large electric fields and DFACs presented here is consistent with them being associated with ionospheric low-conductivity regions. A correlation between the electric field and the derivative of the magnetic field could also be the result of a partially reflected Alfvén wave, but this would not explain why we only observe this correlation for downward currents, or the preference for night-/wintertime conditions. The correlation is also not consistent with the signatures of a U-shaped potential structure. There, the largest current is associated with the centre of the structure, where the perpendicular electric field has its minimum. In fact, in order for the electric field correlation with the DFAC to map all the way out to Cluster altitudes, we must assume that there is no field-aligned potential drop along the magnetic field line. In that case the correlation represents the naked high-altitude signature of the ionospheric density depletion. In many cases we would expect large DFACs to be associated with such a parallel potential drop [e.g. Elphic et al, 1998]; this may be one of the reasons why events of the type we have presented here are relatively rare; we will only see them before such a potential drop has developed. Another reason could be that generally rather low background conductivities will be required.

Reversing the argument, observations of large perpendicular electric fields at magnetospheric altitudes is generally taken as proof that there is a parallel potential drop above the ionosphere. Our results show that this is not necessarily true, but that at least part of this potential drop may be situated deep in the ionosphere, in the E and lower F regions, where the currents partially close through the developing density cavity [Karlsson and Marklund, 1998]. This should be taken into account when interpreting high-altitude electric field data.

For the first event, the current system is stable for around 200 s. The close to linear evolution of $k_{down}$, can be seen as a first observational comparison with modeling of the temporal evolution of ionospheric density cavities. The 200 s time scale is, according to the modeling work quoted above, a typical time scale for creating a deep ionospheric plasma depletion. We would expect to see this type of events for conditions of some moderate geomagnetic activity (to create large DFACs), but not during e.g. the substorm expansion phase, where the current systems would probably move around too much on time scales faster than the depletion time. We have checked this by inspecting the Auroral Electrojet index for the 23 events. Only four of the events where encountered during the expansion phase, whereas the rest were observed during periods that had a medium level of activity; growth or recovery phase or steady magnetospheric convection events. This is further support for the model presented above.

**Acknowledgements**

The authors are grateful to G. Haerendel for some suggestions and comments.

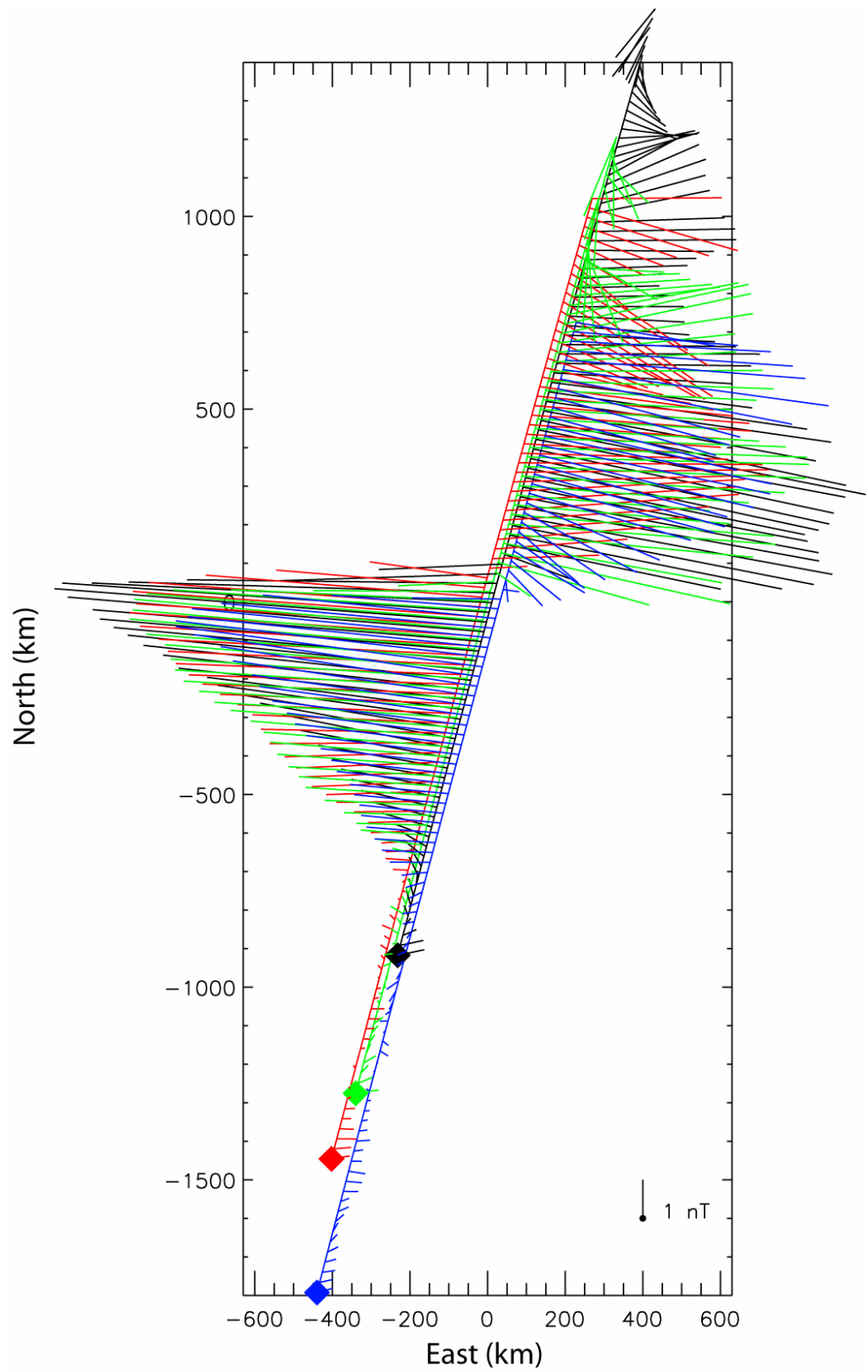

**Figure 1.** Magnetic field perpendicular to the geomagnetic field for the time period 2004-02-18 08:58:20-09:10:00 UT.

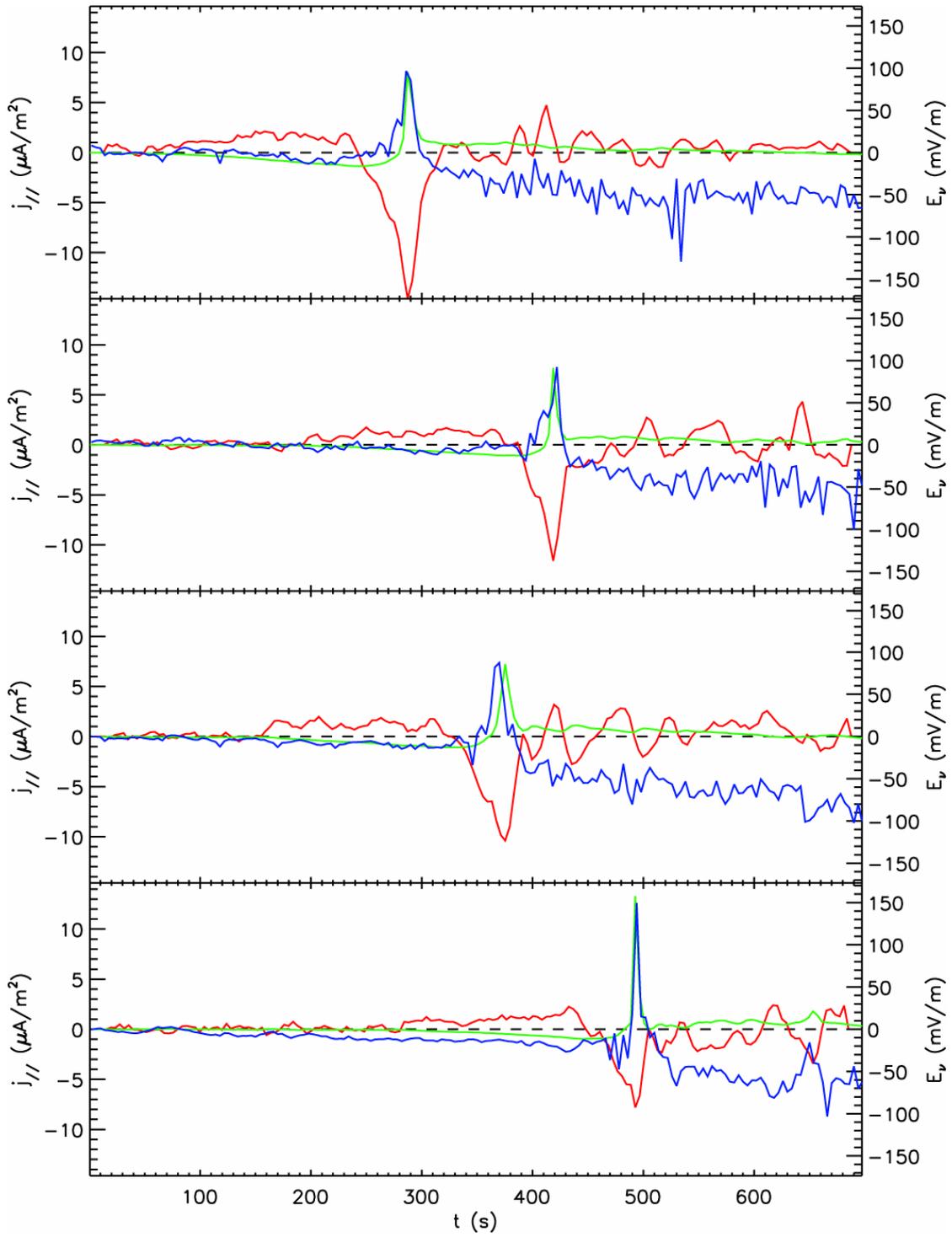

**Figure 2.** Measured (blue) and modeled (green) normal electric field, and measured field-aligned current (red) for the same time period as Figure 1. Note that in this figure downward current is plotted as negative current.

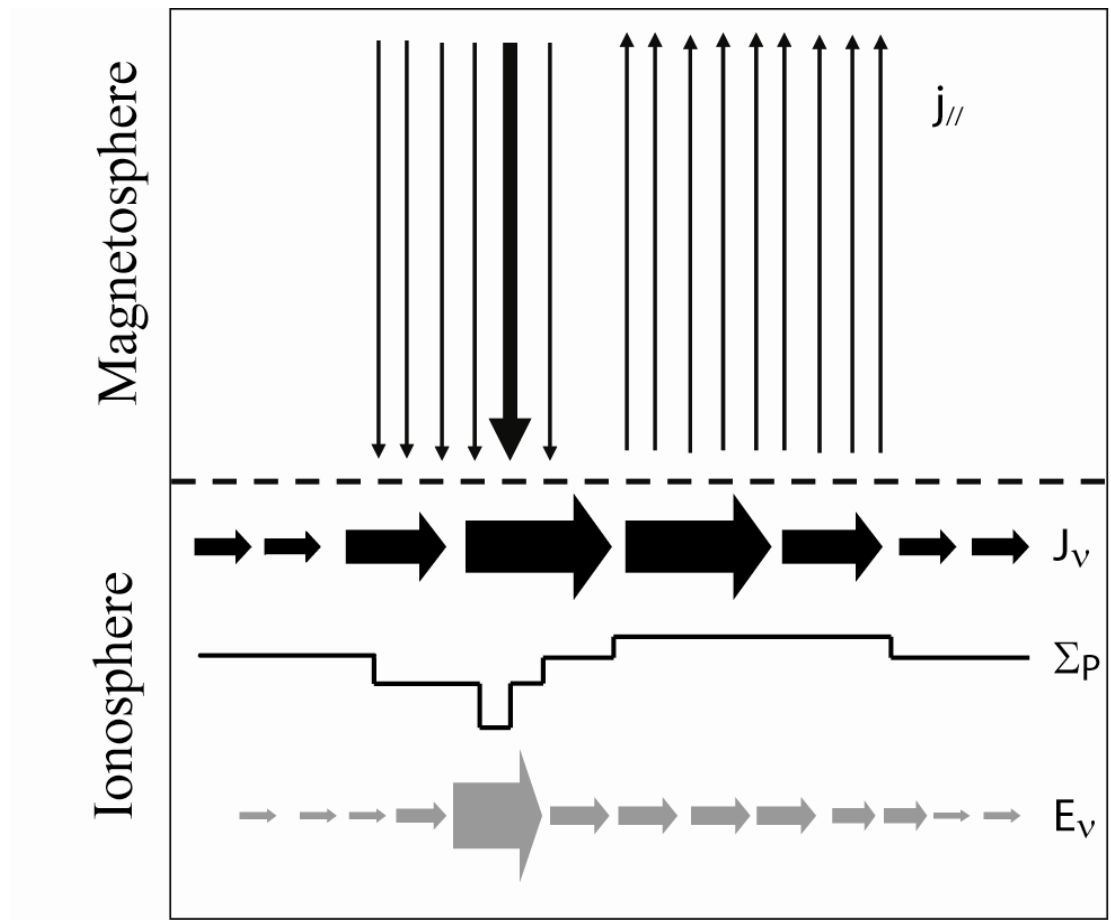

**Figure 3.** Schematic of the connection between DFAC, current closure and ionospheric cavity formation.

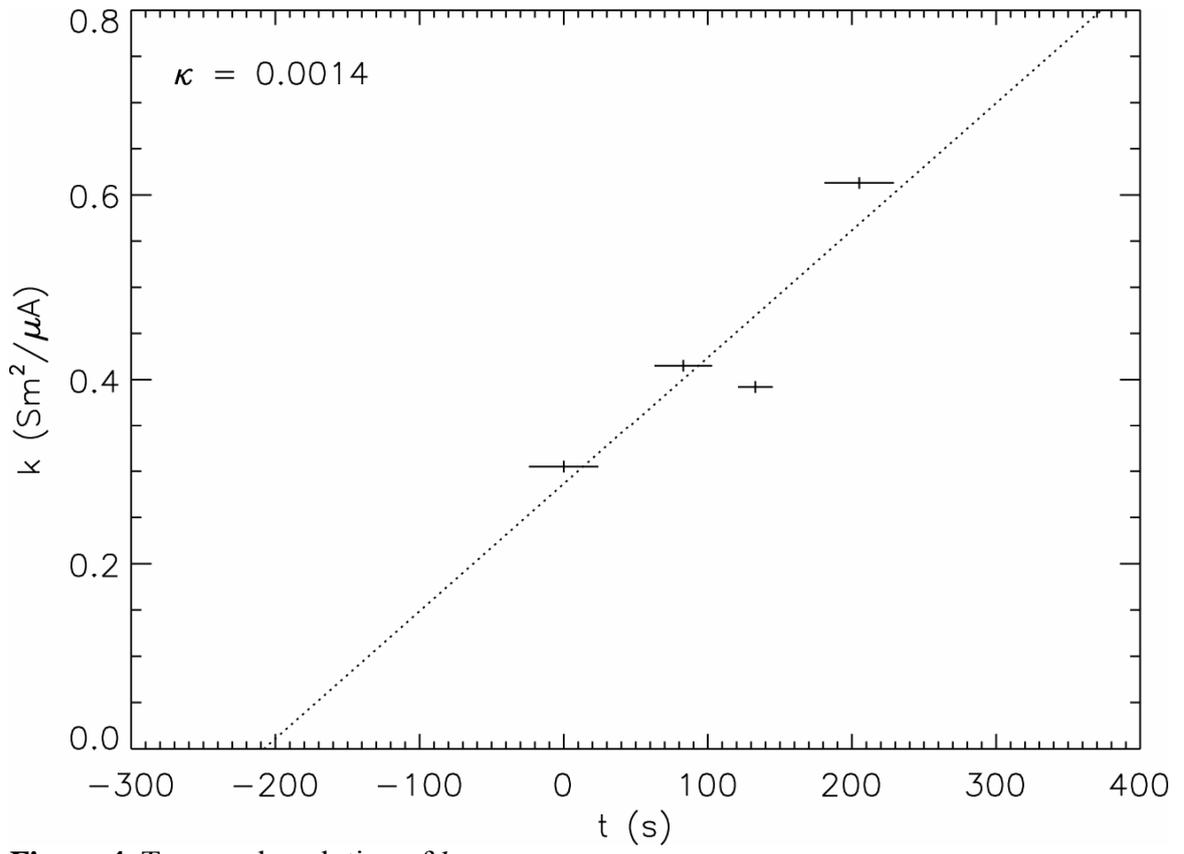

**Figure 4.** Temporal evolution of $k_{down}$.

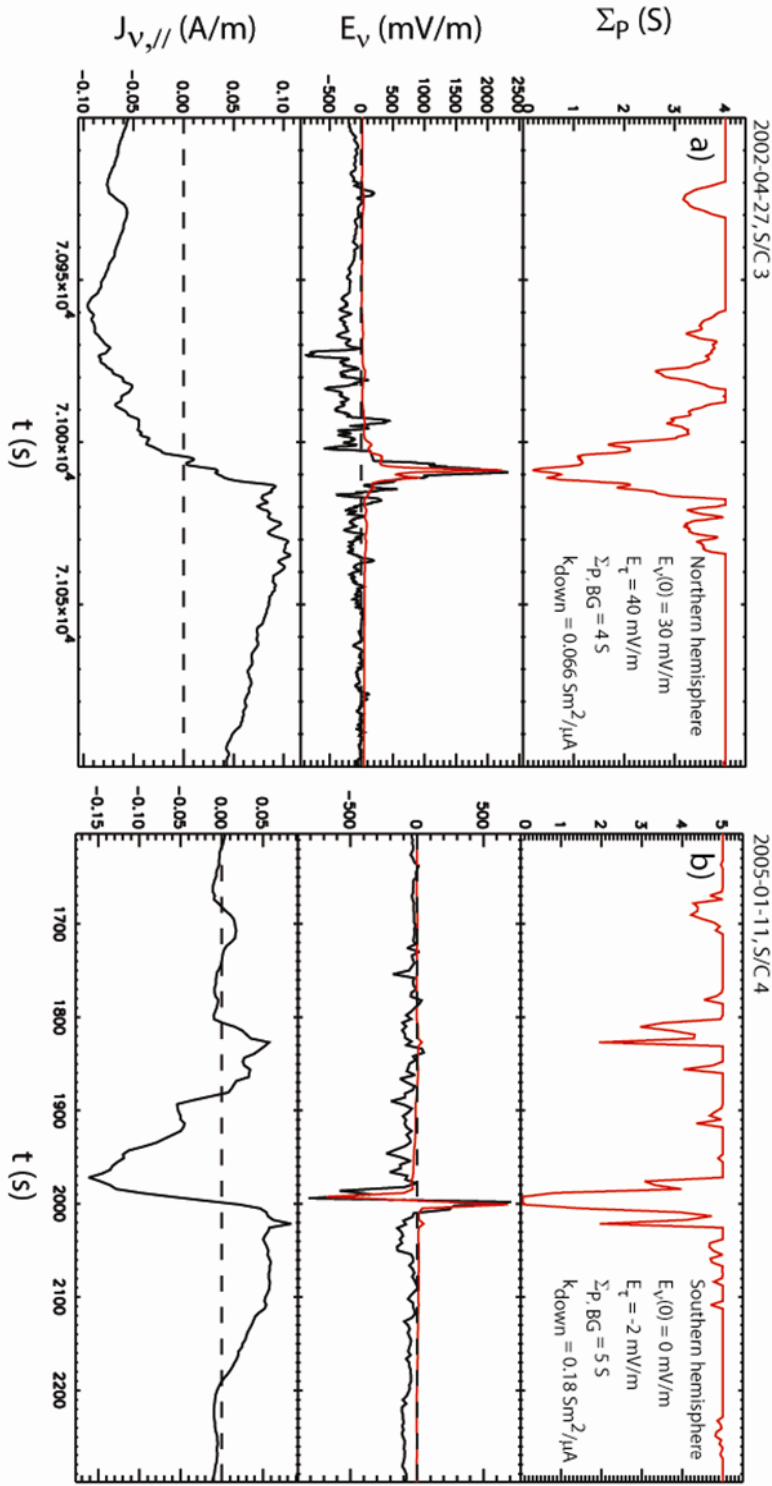

**Figure 5.** Measured (black) and modeled (red) quantities. Time is in seconds from 00:00:00 UT.